\documentclass[12pt]{article}
\usepackage{amssymb}
\usepackage{amsmath}
\usepackage{hyperref}
\usepackage{authblk}
\usepackage{graphicx}
\usepackage{subfig} 
\begin{document}
\title{\bf Effects of the external string cloud on the Van der Waals like behavior and efficiency of AdS-Schwarzschild black holes in massive gravity}

\author[a]{M. Ghanaatian\thanks{Corresponding author: Email:m\_ghanaatian@pnu.ac.ir}}
\author[b]{Mehdi Sadeghi\thanks{ Email:mehdi.sadeghi@abru.ac.ir}}
\author[a]{Hadi Ranjbari\thanks{Email:fhranji@gmail.com}}
\author[a]{Gh. Forozani\thanks{ Email:forozani@pnu.ac.ir}}

\affil[a]{\small{\em{Department of  Physics, Payame Noor University (PNU), P.O.Box 19395-3697  Tehran, Iran}}}
\affil[b]{\small{\em{Department of Physics, School of  Sciences, Ayatollah Boroujerdi University, Boroujerd, Iran}}}

\date{\today}
\maketitle
\abstract{In this paper, we study AdS-Schwarzschild black holes in four and five dimensions in dRGT minimally coupled to a cloud of strings. It is observed  that the entropy of the string cloud and massive terms does not affect the black hole entropy. The observations about four dimensions indicate that the massive term in the presence of external string cloud can not exhibit Van der Waals like behavior for AdS-Schwarzschild black holes and therefore there is only the Hawking-Page phase transition. In contrast,  in five dimensions, the graviton mass modifies this behavior through the third massive term, so that a critical behavior and second order phase transition is deduced. Also, the Joule-Thomson effect is not observed. The black hole stability conditions are also studied in four and five dimensions and a critical value for the string cloud parameter is presented. In five dimensions a degeneracy between states for extremal black holes is investigated. After studying black holes as thermodynamic systems, we consider such systems as heat engines, and finally the efficiency of them is calculated.}\\

\section{Introduction} \label{intro}
In general theory of relativity introduced by Albert Einstein,  graviton is massless. In recent years, the idea of giving mass to the graviton increasingly being considered among cosmologists because they want to reveal the hidden angles of controversial and sophisticated phenomena such as dark energy and dark matter by modifying General Relativity(GR) in order to explain the acceleration of the universe and other unresolved problems in cosmology. The first attempt for constructing a massive theory of gravity is attributed to Fierz and Pauli \cite{Fierz:1939ix} which was done in the context of a linear theory of gravity. Due to some issues of propagator in the massive gravity in the limit $m=0$, it does not reduce to general relativity. Afterwards, people tend to nonlinear massive gravity theories. One of the first person that employed a nonlinear model for massive gravity was Vainshtein \cite{Vainshtein:1972sx}. Accordingly, at some distance below the so-called Vainshtein radius, the linear regime breaks down and the model enters into a nonlinear framework. On the basis of Vainshtein mechanism, the recovery of GR has been established around massive bodies, more details are provided in \cite{DeFelice:2015sya}-\cite{Arraut:2015zga}. But, regrettably, problem of generalization to nonlinear model by Boulware and Deser(BD) suffered from ghosts \cite{Isham:1971gm},\cite{Boulware:1973my}. Nevertheless, recently the BD ghost problem was resolved in \cite{deRham:2010ik},\cite{deRham:2010kj},\cite{Hassan:2011hr} by a nonlinear massive gravity. In these theories, a fixed reference metric can be considered on which the massive gravity propagates. Dynamics of this reference metric is described in the background of theories at present called bi-gravities \cite{Hassan:2011zd},\cite{Hinterbichler:2012cn},\cite{Nomura:2012xr}. Also, higher dimensional nonlinear massive gravity and higher dimensional massive bigravity are investigated in Ref\cite{Do:2016abo} and Ref\cite{Do:2016uef} respectively. The phenomenology of massive gravity has also been interesting to scientists. To review empirical observations in the context of massive gravity see \cite{Gruzinov:2001hp},\cite{Goldhaber:2008xy},\cite{deRham:2016nuf} and the recent LIGO results are given in \cite{TheLIGOScientific:2016src},\cite{Abbott:2016blz}.\\
Black holes are one of the interesting predictions of Einstein theory(GR). The manifestation of black holes as thermodynamical systems have uncovered many aspects of them. As a forerunner of this category, it can be named  the worthwhile work of Hawking and Page who discovered a phase transition between AdS black holes and a global AdS space \cite{Hawking1983}. Then Chamblin et al found a Van der Waals like phase transition in Reissner-Nordstrom AdS black hole \cite{Chamblin:1999tk},\cite{Chamblin:1999hg},\cite{Kastor:2009wy} . As well as Kubiznak and Mann illustrated an interesting analogy between Reissner-Nordstrom AdS black holes and Van der Waals fluid-gas systems in the extended phase space of thermodynamics \cite{Kubiznak:2012wp} . If the first law of black hole is corrected by a VdP term and the cosmological constant is treated as thermodynamical pressure of the black hole and its conjugate variable is regarded as a volume covered by the event horizon of the black hole, then the extended phase space is deduced \cite{Dolan:2010ha},\cite{Johnson:2014yja}. In this illustration, the gravitational mass is regarded as enthalpy.\\
 Another amazing application associated with the introduction of a mechanical work term on the $P-V$ plane is the possibility of considering the black holes as heat engines. In an identified thermodynamic condition, which is determined by the equation of state, it is possible that the black hole burns some substance as fuel and produces mechanical work similar to heat engines. This idea was first introduced by C. Johnson \cite{Johnson:2014yja} . Effects of a string cloud on the criticality and efficiency of AdS black holes as heat engines in the context of GR and f(R) gravity is investigated in \cite{MoraisGraca:2018ofn}. Several authors also have studied black holes as heat engines in some modified theories of gravity \cite{Mo:2018hav}-\cite{Xu:2017ahm}. \\
On the other side, gravity is the low-energy limit of string theory. String theory is a promising theory for the unification of all known forces of nature that interprets particles as vibration modes of one dimensional string objects \cite{Becker:2007zj}. Letelier proposed a model for a cloud of strings, an aggregation of one dimensional objects in a defined geometrical frame that study the gravitational effects of matter such as black holes \cite{Letelier:1979ej}. A cloud of strings is analogous to a pressureless perfect fluid. We are interested to study the effects of the cloud of strings on massive gravitational theory. Many authors have studied various gravitational models with different sources encompassed by a cloud of strings \cite{Ghosh:2014dqa}-\cite{Ganguly:2014cqa}. The impact  of the cloud of strings on Schwarzschild AdS black hole was investigated and its thermodynamical properties in a non-extended phase space was introduced in \cite{Dey:2017xty}. A new extended phase space of Schwarzschild AdS black hole with an energy-momentum tensor coming from a cloud of strings related to the topological charge was illustrated by two formal approaches in \cite{Ghaffarnejad:2018gbf} which leads to the same result in \cite{Dey:2017xty}. It is seen that the effect  of  string  cloud can bring Van der Waals-like behavior and second order phase transition in an extended phase space, while we know Schwarzchild black hole can not display a phase transition.\\
 In this paper we study the thermodynamics of massive-AdS black holes minimally coupled to a cloud of strings in an extended phase space in four and five dimensions and we investigate criticality of these black holes. Our motivation is to simultaneously investigate the effect of string cloud and massive terms on the criticality. This paper is organized as follows. In section 2, the solutions of massive-AdS black holes minimally coupled to a cloud of strings in four dimensions are introduced and the metric function and its diagrams in different modes are investigated. Also, we study the first law of  thermodynamics of these black holes and employ extended phase space to examine the existence of the probable critical points. Then, we  check the behavior of temperature with respect to the radius and seek the stability of black hole with calculating heat capacity. In section 3, we examine the first law of  thermodynamics of these black holes and employ extended phase space thermodynamics to explore critical points in five dimensions. Also, we study the behavior of system along the coexistence line by plotting isothermal curves in $P - T$ diagrams. We then investigate the possibility of the Joule-Thomson effect \cite{Okcu:2016tgt} in our model by drawing isenthalpic curves in $T - P$ plan. In the following, the critical exponents are calculated  in our model. The thermal stability of the solutions in canonical ensemble are studied. Then, in section 4, we consider black holes as heat engines and we calculate the efficiency of them. Finally, in section 5, we present our work results.
\\

\section{Gravity setup and thermodynamics in $d=4$ dimensions  }
 \label{sec2}
The action of GR-$\Lambda $-massive gravity coupled to a cloud of strings is 
\begin{equation}
I =  - \frac{1}{{16\pi }}\int {{d^d}x\sqrt { - g} [R - 2\Lambda  + {m^2}\sum\limits_{i = 1}^4 {{c_i}{{\cal U}_i}(g,f)} ]}  + \int_\Sigma  {{{\rm N}_{\rm P}}} \sqrt { - \chi } d{\lambda ^0}d{\lambda ^1},
\end{equation}
where $R$ is the scalar curvature of the metric ${g_{\mu \upsilon }}$, $\Lambda  = {{ - n(n - 1)} \over {2{l^2}}}$ is the negative cosmological constant $(n = d - 1)$ with $ l$ as the cosmological constant scale,  $m$ is massive parameter and  ${f_{\mu \upsilon }}$ is fixed symmetric tensor. The ${{c_i}}$'s are constant and the ${{{\cal U}_i}}$'s are symmetric polynomials of the eigenvalues of the d$ \times $d matrix ${\cal K}_\upsilon ^\mu  = \sqrt {{g^{\mu \alpha }}{f_{\alpha \upsilon }}} $, where 
 \begin{align}\label{U}
  & \mathcal{U}_1=[\mathcal{K}]\nonumber,\\
  & \mathcal{U}_2=[\mathcal{K}]^2-[\mathcal{K}^2]\nonumber,\\
  &\mathcal{U}_3=[\mathcal{K}]^3-3[\mathcal{K}][\mathcal{K}^2]+2[\mathcal{K}^3]\nonumber,\\
  & \mathcal{U}_4=[\mathcal{K}]^4-6[\mathcal{K}^2][\mathcal{K}]^2+8[\mathcal{K}^3][\mathcal{K}]+3[\mathcal{K}^2]^2-6[\mathcal{K}^4].
   \end{align}
The square root in ${\cal K}$ means $(\sqrt {\rm A} )_\upsilon ^\mu (\sqrt {\rm A} )_\lambda ^\upsilon  = {\rm A}_\lambda ^\mu $ and the rectangular brackets denote traces.
The second integral called a Nambu-Goto action. (${\lambda ^0}$,${\lambda ^1}$) is a parametrization of the world sheet $\Sigma $ and ${{\rm N}_{\rm P}}$  is positive and is related to the tension of the string. $\chi $ is the determinant of the induced metric 
\begin{equation}
{\chi _{ab}} = {g_{\mu \upsilon }}\frac{{\partial {x^\mu }}}{{\partial {\lambda ^a}}}\frac{{\partial {x^\upsilon }}}{{\partial {\lambda ^b}}}.
\end{equation}
The action can also be described by a spacetime bi-vector ${\Sigma ^{\mu \upsilon }}$, given by 
\begin{equation}
{\Sigma ^{\mu \upsilon }} = {\varepsilon ^{ab}}\frac{{\partial {x^\mu }}}{{\partial {\lambda ^a}}}\frac{{\partial {x^\upsilon }}}{{\partial {\lambda ^b}}},
\end{equation}
in which ${\varepsilon ^{ab}}$ is Levi-Civita tensor.\\
The Nambu-Goto action can be written as 
\begin{equation}
{I_{NG}} = {{\rm N}_{\rm P}}\int\limits_\Sigma  {\sqrt { - \frac{1}{2}{\Sigma _{\mu \upsilon }}{\Sigma ^{\mu \upsilon }}} } d{\lambda ^0}d{\lambda ^1}.
\end{equation}
The energy-momentum tensor for the string can be calculated from the relation ${{\rm T}_{\mu \upsilon }} = {{ - 2\partial {\cal L}} \over {\partial {g^{\mu \upsilon }}}}$, where ${\cal L} = {{\rm N}_{\rm P}}\sqrt { - {1 \over 2}{\Sigma _{\mu \upsilon }}{\Sigma ^{\mu \upsilon }}} $ . The energy-momentum tensor of the string cloud is then given by 
\begin{equation}
{{\rm T}^{\mu \upsilon }} = \rho \frac{{{\Sigma ^{\mu \sigma }}\Sigma _\sigma ^\upsilon }}{{\sqrt { - \chi } }},
\end{equation}
where $\rho $  is the density of the string cloud.\\
The equation of motion is as follows,
\begin{equation}\label{EOM}
{G_{\mu \upsilon }} + \Lambda {g_{\mu \upsilon }} + {m^2}{X_{\mu \upsilon }} = {{\rm T}_{\mu \upsilon }},
\end{equation}
in which ${G_{\mu \upsilon }}$  is the Einstein tensor and  ${X_{\mu \upsilon }}$  is
\begin{align*}\label{8}
{X _{\mu \upsilon }} = \frac{{{c_1}}}{2}({{\cal K}_{\mu \upsilon }} - {{\cal U}_1}{g_{\mu \upsilon }}) - \frac{{{c_2}}}{2}({{\cal U}_2}{g_{\mu \upsilon }} - 2{{\cal U}_1}{{\cal K}_{\mu \upsilon }} + 2{\cal K}_{\mu \upsilon }^2)\\
 - \frac{{{c_3}}}{2}({{\cal U}_3}{g_{\mu \upsilon }} - 3{{\cal U}_2}{{\cal K}_{\mu \upsilon }} + 6{{\cal U}_1}{\cal K}_{\mu \upsilon }^2 - 6{\cal K}_{\mu \upsilon }^3) - \frac{{{c_4}}}{2} \times \\
({{\cal U}_4}{g_{\mu \upsilon }} - 4{{\cal U}_3}{{\cal K}_{\mu \upsilon }} + 12{{\cal U}_2}{\cal K}_{\mu \upsilon }^2 - 24{{\cal U}_1}{\cal K}_{\mu \upsilon }^3 + 24{\cal K}_{\mu \upsilon }^4),
\end{align*}
 and  conservation of the energy-momentum tensor, ${\nabla _\upsilon }{{\rm T}^{\mu \upsilon }} = 0$  leads
\begin{equation}
{\partial _\mu }(\sqrt { - g} \rho {\Sigma ^{\mu \sigma }}) = 0.
\end{equation}
$ f_{\mu \nu} $ was proposed in \cite{Sadeghi:2018vrf} with the  form
	$ f_{\mu \nu} = \frac{c_0^2}{l^2}diag(0,0,1,1)$.
Then considering this ansatz for the metric,
\begin{equation}
d{s^2} =  - f(r)d{t^2} + {f^{ - 1}}(r)d{r^2} + {r^2}{h_{ij}}d{x_i}d{x_j},\\
\end{equation}
yielding \cite{Sadeghi:2019}
 \begin{align}\label{U_4} 
 & \mathcal{U}_1=\frac{3c_{0}}{r}, \,\,\,  \,\,\, \mathcal{U}_2=\frac{6c_0^2}{r^2},\,\,\,\,\mathcal{U}_3=0,\,\,\,\,\mathcal{U}_4=0.\nonumber
 \end{align}
Finally, the field equations yield 
\begin{equation}
f(r) = k - \frac{b}{r} - a - \frac{\Lambda }{3}{r^2} + {m^2}(\frac{{{c_0}{c_1}}}{2}r + c_0^2{c_2}),
\end{equation}
where $b$ is an integration constant which is calculated from solving $f({r_0}) = 0$  that ${r_0}$  is location of the event horizon  
\begin{equation}
b = {r_0}\left[ {k - \frac{\Lambda }{3}r_0^2 - a + \Delta } \right],
\end{equation}
\begin{equation}
\Delta  \equiv {m^2}(\frac{{{c_0}{c_1}}}{2}{r_0} + c_0^2{c_2}),
\end{equation}
and $a$ is a positive constant.\\
To investigate the effect of the string cloud on our solution we plot  $f(r) - r$ diagrams (see Figure 1) . We see that based on the value of the string cloud parameter, for $a < {a_c}$ maximum three roots appear, while for $a > {a_c}$ maximum one root appears, where the value of ${a_c}$ is given in the following . We  note that the metric function  has a maximum of three roots regardless of curvature geometry which means, it has a maximum of three event horizon apart from the topology of the horizon: sperical $(k = 1)$, flat $(k = 0)$ or hyperbolic $(k =-1)$  in 4 dimensions.\\
\begin{figure}[ht]
    \centering
\subfloat[A]{\includegraphics[width=4.5cm]{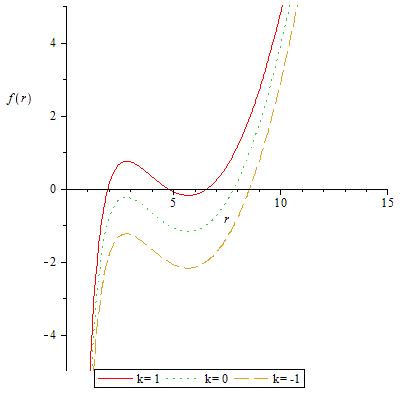}}
\qquad
\subfloat[B]{\includegraphics[width=4.5cm]{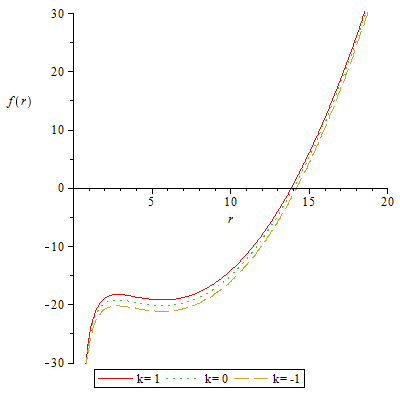}}
\qquad
\subfloat[C]{\includegraphics[width=4.5cm]{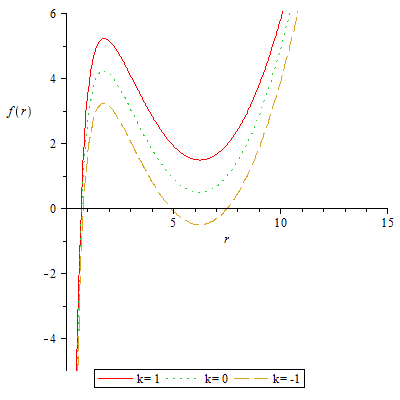}}
\caption{ $f(r) - r$  for $d=4$, $\Lambda  =  - 1$, ${c_0} = 1$, ${c_1} =  - 2$, ${c_2} = 4$, $b = 20$ and $m = 2.1$, (a): $a$=1$< {a_c}$, (b): $a$=20$>{a_c}$ and (c): $b = 10$,  $a$=1$< {a_c}$. }
\end{figure}
In an extended phase space we consider the cosmological constant as pressure of the black hole with $P =  - {\Lambda  \over {8\pi}}$, then its conjugated variable plays role of the black hole's volume. In this conjecture the black hole mass is interpreted as enthalpy \\
\begin{equation}
M = \frac{{{V_2}b}}{{8\pi }},
\end{equation}
where ${{V_2}}$, the area of a unit volume of constant $(t,r)$ space is equal to $4\pi $.\\
Entropy $S$ and temperature $T$ of the black hole can be derived respectively as follow
\begin{equation}
S = \int_0^{{r_0}} {\frac{1}{T}} (\frac{{\partial M}}{{\partial {r_0}}})d{r_0} = \frac{{{V_2}r_0^2}}{4},
 \end{equation}
\begin{equation}
T = \frac{1}{{4\pi }}{\partial _r}f(r)\left| {_{r = {r_0}}} \right. = \frac{k}{{4\pi {r_0}}} - \frac{{{r_0}\Lambda }}{{4\pi }} - \frac{a}{{4\pi {r_0}}} + \frac{{{m^2}}}{{4\pi {r_0}}}({c_0}{c_1}{r_0} + c_0^2{c_2})].
 \end{equation}
Then we find that these thermodynamic quantities satisfy the first law of the black hole thermodynamics in the extended phase space with the following form \\
\begin{equation}
dM = TdS + VdP+ Ada  + {C_1}d{c_1} + {C_2}d{c_2},
 \end{equation}
with 
\begin{equation}
T = {(\frac{{\partial M}}{{\partial S}})_{a,P,{c_i}}},
\end{equation}
\begin{equation}
V = {(\frac{{\partial M}}{{\partial P}})_{S,a,{c_i}}} = \frac{1}{3}{V_2}r_0^3,
\end{equation}
\begin{equation}
A = {(\frac{{\partial M}}{{\partial a}})_{S,P,{c_i}}} =  - \frac{{{V_2}}}{{8\pi }}{r_0},
\end{equation}
\begin{equation}
{C_1} = {(\frac{{\partial M}}{{\partial {c_1}}})_{S,a,P,{c_2},{c_3}}} = \frac{{{V_2}{m^2}{c_0}r_0^2}}{{16\pi }},
\end{equation}
\begin{equation}
{C_2} = {(\frac{{\partial M}}{{\partial {c_2}}})_{S,a,P,{c_1},{c_3}}} = \frac{{{V_2}{m^2}c_0^2{r_0}}}{{8\pi }}.
\end{equation}
where $V$ is the thermodynamic volume, $A$ and ${C_i}$'s stand for the physical quantities conjugated to the parameters $a$ and ${c_i}$'s respectively.
Besides, the corresponding Smarr relation can be extracted by a scaling argument as
\begin{equation}
M = 2TS - 2VP - {C_1}{c_1}.
 \end{equation}
Since the ${{c_2}}$-term and $a$-term in the metric function are constant terms in four dimensions do not appear in first law of black hole thermodynamics and we set $d{c_2} = 0,da = 0$. Also, if we define an equipotential surface $f(r) = cte$ and varying it with respect to variables $k$, $r$, ${m_0} \equiv b/2$, $a$ and ${c_i}$, then a term $\omega d\varepsilon $ is added to equation (16) where $\varepsilon  = {V_2}k$ is named topological charge  and $\omega  = {{{k^0}r} \over {8\pi }}$ is its conjugated potential. Anyway since the dimention of $\varepsilon $ is proportional to ${[L]^0}$ this term has no role in coresponding Smarr relation.\\
By inserting relationships (11),(13),(14),(18) and (20) in (22) we will achieve  the equation of state 
\begin{equation}
P = \frac{1}{{2{r_0}}}(T - \frac{{{m^2}{c_0}{c_1}}}{{4{\pi _{}}}}) + \frac{{a - k - {m^2}c_0^2{c_2}}}{{8\pi r_0^2}},
 \end{equation}
where the value of $( - {{{m^2}{c_0}{c_1}} \over {4\pi }})$ can be considered as a correction to the Hawking temperature, which is imposed by massive graviton.\\
The ideal gas law for one mole of gas composed of non-interacting point particles  satisfy $P \propto {T \over V}$. Van der Waals proposed that  all particles are hard spheres of the same finite radius r (the van der Waals radius). The available space in which the particles are free to move  is limited by the amount of space occupied by themselves, so  Van der Waals  corrected the gas state equation by replacing $V - B$ instead $V$, where $B$ is called the co-volume. Then we will have $P \propto {T \over {V -B}} = T\left( {{1 \over V} + {B \over {{V^2}}} + {{{B^2}} \over {{V^3}}} + ...} \right)$. Consequently in order to the equation of state display Van der Waals like behaviour,as a necessary condition, it should include at least the ${1 \over V}\left( { \propto {1 \over {r_0^3}}} \right)$ term and the higher powers in the denominator. Thus, in our case study, the equation of state (23), does not show the  Van der Waals like behavior.\\
Also we can explore critical phenomena and van der Waals like behaviour by computing the inflection point ${({{\partial P} \over {\partial {r_0}}})_T} = {({{{\partial ^2}P} \over {\partial {r_0}^2}})_T} = 0$,  we see that no critical points are found. Therefore we observe that adding the graviton mass to the AdS-Schwarzschild black holes coupled to a cloud of strings in 4 dimensions have no critical behaviour. This fact can also be seen from the $P -{r_0}$ diagrams(see Figure 2). Here, as in the case of  AdS-Schwarzschild black hole, there is only the Hawking-Page phase transition. This is evident from the $T-{r_0}$ diagrams(see Figure 2). These diagrams are plotted for two conditions: $a > {a_c}$ and $a < {a_c}$ where ${a_c} =k+ {m^2}c_0^2{c_2}  $. For  $a < {a_c}$, we see that there is a minimum temperature in which a phase transition  between a small black hole and a large black hole takes place.\\
To further explore thermodynamical properties of the black holes, we calculate the specific heat in canonical ensemble to check the stability of the black holes. The specific heat can be derived as\\
\begin{equation}
C = \frac{{\partial M}}{{\partial T}} = \frac{{2\pi r_0^2( - \Lambda r_0^2 + k - a + {m^2}({c_0}{c_1}r_0^{} + c_0^2{c_2}))}}{{ - \Lambda r_0^2 - ({m^2}c_0^2{c_2} + k - a)}}.
 \end{equation}
The $C-{r_0}$  diagrams  are plotted in Figure 3. We see that for $a < {a_c}$   the black holes with radius from ${r_0} = {1 \over {2\left| \Lambda  \right|}}[ - {m^2}{c_0}{c_1} - \sqrt {{m^4}c_0^2c_1^2 - 4\left| \Lambda  \right|({m^2}c_0^2{c_2} + k - a)} ]$ to ${r_0} = \sqrt {{{{m^2}c_0^2{c_2} + k - a} \over {\left| \Lambda  \right|}}} $ and ${r_0} > {1 \over {2\left| \Lambda  \right|}}[ - {m^2}{c_0}{c_1} + \sqrt {{m^4}c_0^2c_1^2 - 4\left| \Lambda  \right|({m^2}c_0^2{c_2} + k - a)} ]$  are stable and for the rest of the radius are unstable. The diagram for $a > {a_c}$ shows that the black hole is stable .\\
It should be noted that, as can be deduced from reference \cite{Parvizi:2017boc}, the physical condition of unitarity is established. Also, considering that the dominent term near the boundary is the first massive term, causality requires that ${c_1} \le 0$. Therefore, only those solutions that satisfy this condition are physical. If ${c_1} = 0$ then ${c_2} \le 0$ and so on.\\  
The Gibbs free energy for $d = 4$  is given by
 \begin{equation}
G = M - TS = - \frac{1}{4}(a - k - {m^2}c_0^2{c_2}){r_0} - \frac{{2\pi }}{3}P{r_0}^3.
 \end{equation}
The $G-T$ diagrams for $d=4$ are plotted in Figure 4. It is known that in the second order phase transition, due to a thermodynamic potential such as Gibbs free energy becomes non-analytic,  a swallow tail-like behaviour and discontinuity appears in $G-T$ diagrams, which is not observed here.
\begin{figure}[ht]
    \centering
\subfloat[a]{\includegraphics[width=4.5cm]{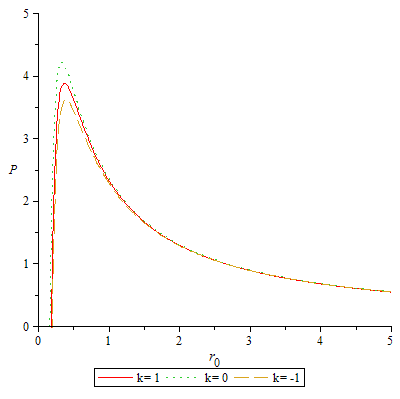}}
\qquad
\subfloat[b]{\includegraphics[width=4.5cm]{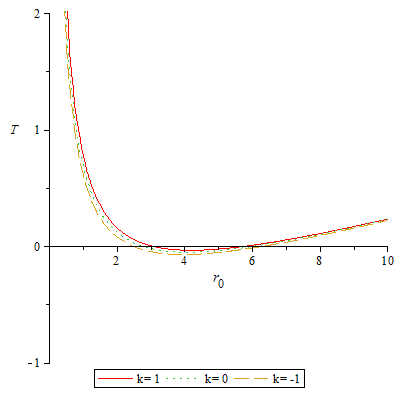}}
\qquad
\subfloat[c]{\includegraphics[width=4.5cm]{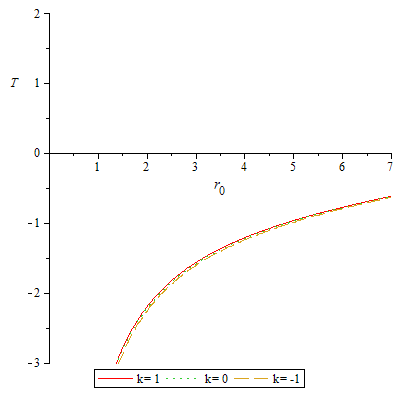}}
\caption{$(a)$: $P -{r_0}$  for $d=4$, $T=5$, ${c_0} = 1$, ${c_1} =  - 2$, ${c_2} = 3.18$, $m = 2.1$ and $a=1$, $(b)$: $T -{r_0}$  for  $d=4$, $\Lambda  =  - 1$, ${c_0} = 1$, ${c_1} =  - 2$, ${c_2} = 3.18$, $m = 2.1$, $a=1<{a_c}$ , $(c)$: $a=60>{a_c}$. }
\end{figure}
\begin{figure}[ht]
    \centering
\subfloat[A]{\includegraphics[width=5cm]{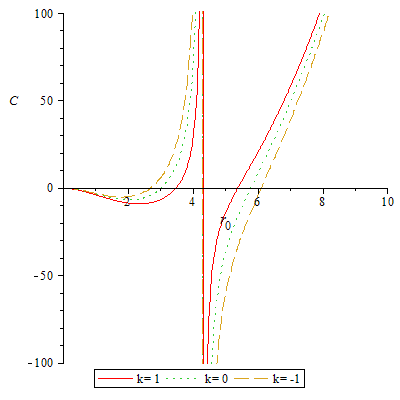}}
\qquad
\subfloat[B]{\includegraphics[width=5cm]{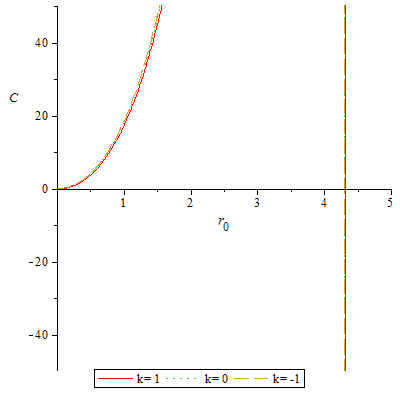}}
\caption{ $C-{r_0}$  for $d=4$, $\Lambda  = -1$, ${c_0} = 1$, ${c_1} = -2$,  ${c_2} = 3.18$, and $m = 2.1$ , (a):$a=1$ and (b): $a=60$.}
\end{figure}
\begin{figure}[ht]
    \centering
\subfloat[A]{\includegraphics[width=5cm]{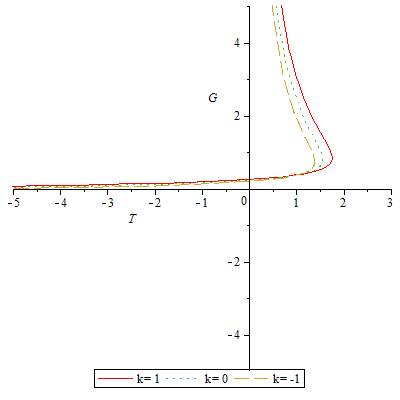}}
\qquad
\subfloat[B]{\includegraphics[width=5cm]{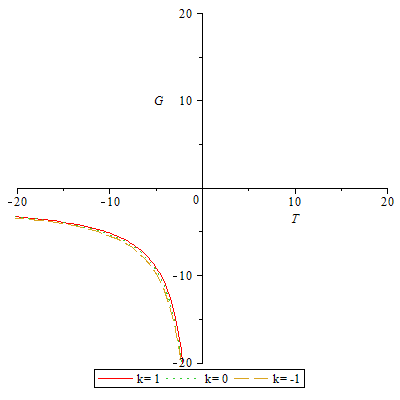}}
\caption{ $G-T$  for $d=4$, $P=1$, ${c_0} = 1$, ${c_1} = -2$, ${c_2} = 3.18$, and $m = 2.1$, (a): $a$=1 and (b): $a$=60.}
\end{figure}

\section{Critical behavior in $d =5$ dimensions}
In five dimensions, $\mathcal{U}_i$'s are the following \cite{Sadeghi:2015vaa},\cite{Sadeghi:2018ylh}
 \begin{align}\label{U_5} 
 & \mathcal{U}_1=\frac{3c_{0}}{r}, \,\,\,  \,\,\, \mathcal{U}_2=\frac{6c_0^2}{r^2},\,\,\,\,\mathcal{U}_3=\frac{6c_0^3}{r^3},\,\,\,\,\mathcal{U}_4=0,\nonumber
 \end{align} 
therefore the metric function is given by\\
 \begin{equation}
f(r) = k - \frac{b'}{{{r^2}}} - \frac{2a}{{3r}} - \frac{\Lambda }{6}{r^2} + {m^2}(\frac{{{c_0}{c_1}}}{2}r + c_0^2{c_2} + \frac{{c_0^3{c_3}}}{r}),
 \end{equation}
with 
 \begin{equation}
b' = r_0^2\left[ {k - \frac{\Lambda }{6}r_0^2 - \frac{2a}{{3{r_0}}} + {m^2}(\frac{{{c_0}{c_1}}}{2}r_0 + c_0^2{c_2} + \frac{{c_0^3{c_3}}}{r_0}) } \right].
 \end{equation}
$f(r)-r$ diagrams show that  in five dimensions we have at most three  horizons, so the effect of string cloud on the number of event horizons is similar to the four dimensions (see Figure 5).
\begin{figure}[ht]
    \centering
\subfloat[A]{\includegraphics[width=5cm]{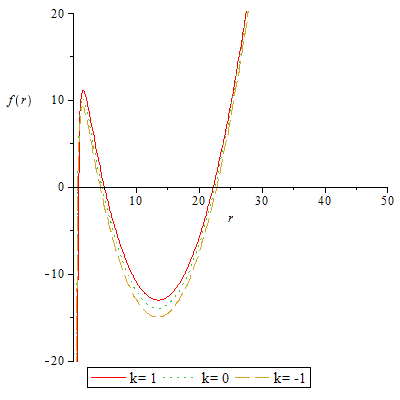}}
\qquad
\subfloat[B]{\includegraphics[width=5cm]{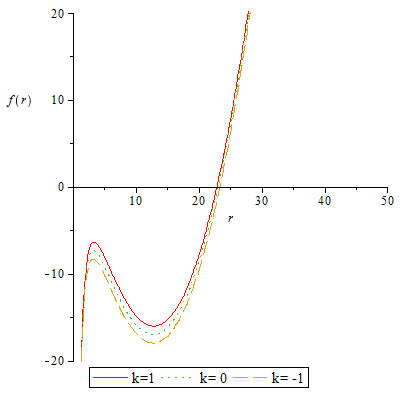}}
\caption{ $f(r) - r$  for $d=5$, $\Lambda  = -1$, ${c_0} = 1$, ${c_1} = -2$, ${c_2} = 3.18$, ${c_3} = 4$, $b' = 20$ and $m = 2.1$, (a): $a$=1$< {a_c}$ and (b): $a$=60$> {a_c}$.}
\end{figure}
The ADM mass $M$, Wald entropy $S$ and Hawking temperature $T$  of the black hole can be calculated respectively as follow
 \begin{equation}
M = \frac{{3{V_3}}}{{16\pi }}r_0^2[k - \frac{2a}{{3{r_0}}} + \frac{{4\pi }}{3}Pr_0^2 + {m^2}(\frac{{{c_0}{c_1}}}{2}r_0 + c_0^2{c_2} + \frac{{c_0^3{c_3}}}{r_0}) ],
 \end{equation}
 \begin{equation}
S = \frac{{{V_3}r_0^3}}{4},
 \end{equation}
 \begin{equation}
T = \frac{k}{{2\pi {r_0}}} - \frac{{{r_0}\Lambda }}{{6\pi }} - \frac{a}{{6\pi r_0^2}} + \frac{{{m^2}}}{{4\pi {r_0}}}(\frac{3}{2}{c_0}{c_1}{r_0} + 2c_0^2{c_2} + \frac{{c_0^3{c_3}}}{{{r_0}}}),
  \end{equation}
where ${{V_3}}$,  volume of the  three dimensional unit sphere as plane or hyperbola, is equal to ${{4\pi } \over 3}$ . These relations satisfy  the first law of black hole thermodynamics in the extended phase space. The corresponding Smarr relation  can be derived as
 \begin{equation}
2M = 3TS - 2VP + Aa - {C_1}{c_1} +{C_3}{c_3},
 \end{equation}
where 
  \begin{equation}
V = {(\frac{{\partial M}}{{\partial P}})_{S,a,{c_i}}} = \frac{1}{4}{V_3}r_0^4,
 \end{equation}
 \begin{equation}
A = {(\frac{{\partial M}}{{\partial a}})_{S,P,{c_i}}} = \frac{{ - {r_0}{V_3}}}{{8\pi }},
 \end{equation}
 \begin{equation}
{C_1} = {(\frac{{\partial M}}{{\partial {c_1}}})_{S,a,P,{c_2},{c_3}}} = \frac{{3{V_3}{m^2}{c_0}r_0^3}}{{32\pi }},
\end{equation}
\begin{equation}
{C_3} = {(\frac{{\partial M}}{{\partial {c_3}}})_{S,a,P,{c_1},{c_2}}} = \frac{{3{V_3}{m^2}c_0^3{r_0}}}{{16\pi }}.
\end{equation}
By inserting these relations in Eq.(31), and using (28) and (29), the pressure is obtained
\begin{equation}
P = \frac{3}{{4{r_{_0}}}}(T - \frac{{3{m^2}{c_0}{c_1}}}{{8\pi }}) - \frac{{3k + 3{m^2}c_0^2{c_2}}}{{8\pi r_0^2}} + \frac{{2a - 3{m^2}c_0^3{c_3}}}{{16\pi r_0^3}}.
\end{equation}
With setting ${({{\partial P} \over {\partial {r_0}}})_T} = {({{{\partial ^2}P} \over {\partial {r_0}^2}})_T} = 0$, the critical radius, temperature and pressure as follow
\begin{equation}
{r_{0c}} = \frac{{2a - 3{m^2}c_0^3{c_3}}}{{2k + 2{m^2}c_0^2{c_2}}},
\end{equation}
 \begin{equation}
{T_c} = \frac{{{{(k + {m^2}c_0^2{c_2})}^2}}}{{\pi (2a - 3{m^2}c_0^3{c_3})}} + \frac{{3{m^2}{c_0}{c_1}}}{{8\pi }},
\end{equation}
 \begin{equation}
{P_c} = \frac{{{{(k + {m^2}c_0^2{c_2})}^3}}}{{2\pi {{(2a - 3{m^2}c_0^3{c_3})}^2}}}.
\end{equation}
The appearance of these values indicates critical behavior in five dimensions for $a < {a_c}$ (see Figure 6). We note that  if we exclude the massive term($m \to 0$), the above equations  are converted to the results in Ref \cite{Ghaffarnejad:2018gbf} which is valid for $k=1$, therefore in the absence of the massive term, we have critical behavior only for spherical topology. In other words Einstein gravity modification with massive graviton in the presence of external string cloud can bring second order phase transition and Van der Waals like behavior for topological black holes($k = 0, - 1$). We can consider the value of $( - {{3{m^2}{c_0}{c_1}} \over {8\pi }})$ for the correction to the Hawking temperature. If the graviton mass corrects Hawking's temperature in this way, we can estimate the compression factor or the gas deviation factor as ${{{P_c}{v_c}} \over {{T_c}}} = {1 \over 3} \simeq {{2.66} \over 8}$, which is different from ${{{P_c}{v_c}} \over {{T_c}}} = {3 \over 8}$ for the Van der Waals fluid, where $v = {4 \over 3}{r_0}$ is defined as specific volume.\\
In critical behavior the conditions of positive pressure and radius are required
\begin{equation}
k + {m^2}c_0^2{c_2} > 0,
\end{equation}
\begin{equation}
2a - 3{m^2}c_0^3{c_3} > 0.
\end{equation}
In Figure 6, the $P - {r_0}$ diagrams are plotted for ${c_3} < 0$. It should be noted that if ${c_3} > 0$, then it is necessary that $a > {3 \over 2}{m^2}c_0^3{c_3}$  to maintain the condition (41).
The condition of positive pressure results
\begin{equation}
T > \frac{{3{m^2}{c_0}{c_1}}}{{8\pi }} + \frac{{2{{(k + {m^2}c_0^2{c_2})}^2}}}{{3\pi (2a - 3{m^2}c_0^3{c_3})}},
\end{equation}
which equation (38) applies in this constraint.\\
\begin{figure}[ht]
    \centering
\subfloat[A]{\includegraphics[width=5cm]{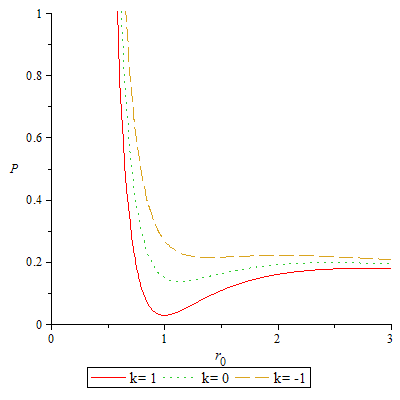}}
\qquad
\subfloat[B]{\includegraphics[width=5cm]{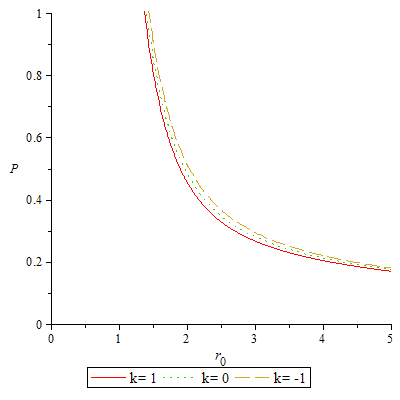}}
\caption{ $P -{r_0}$  for $d=5$, $T=0.5$, ${c_0} = 1$, ${c_1} = -2$, ${c_2} = 4$, ${c_3} = -4$ and $m = 2.1$, (a): $a$=1$< {a_c}$, and (b): $a$=60$> {a_c}$.}
\end{figure}
We also can study the critical behavior of black hole temperature in $P - r$ plane and investigate the effect of cloud string parameter on that (Figure 7). We see that when the temperature is above the critical value, a behavior similar to that of an ideal gas appears, which is referred to as the ideal gas phase transition. But for temperatures below the critical temperature, three branches are seen, representing small, medium, and large black holes. Except for the latter, which is unstable, the other two are stable and consistent with the Van der Waals liquid/gas phase transition. As seen in Figure 7-(b), there exists a particular temperature
\begin{equation}
{T_ \bot } = \frac{3}{{4\pi }}\frac{{{{\left( {k + {m^2}c_0^2{c_2}} \right)}^2}}}{{\left( {2a - 3{m^2}c_0^3{c_3}} \right)}} + \frac{{3{m^2}{c_0}{c_1}}}{{8\pi }},
 \end{equation}
for which we have $\partial P/\left( {\partial r_0{\rm{ }}} \right)\left| {_{T = {T_ \bot }}} \right. = P\left| {_{T = {T_ \bot }}} \right. = 0$, similar to what we have seen in Van der Waals fluid.\\
\begin{figure}[ht]
    \centering
\subfloat[A]{\includegraphics[width=5cm]{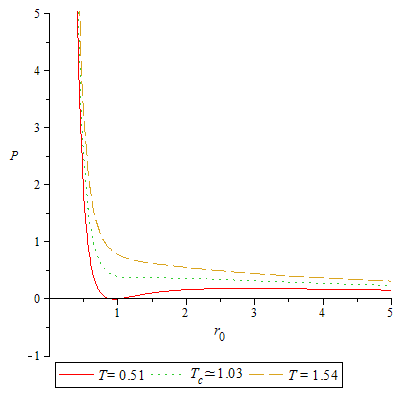}}
\qquad
\subfloat[B]{\includegraphics[width=5cm]{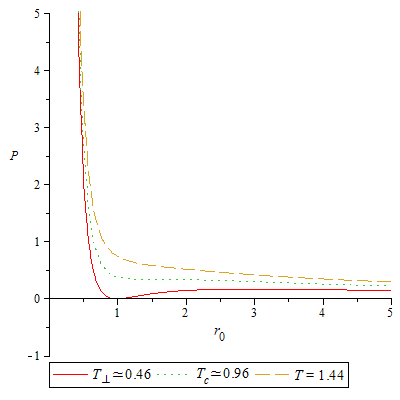}}
\caption{ $P -{r_0}$  for $d=5$, $k = 1$, ${c_0} = 1$, ${c_1} = -2$, ${c_2} = 4$, ${c_3} = -4$ and $m = 2.1$, (a): $a$=0.1 and (b): $a$=1.}
\end{figure}
An interesting way to analyze the phase transition is to plot the $P - T$ diagram for two different phases where the black hole phase transition is between the two so that both phases have the same Gibbs free energy. This phase transition is of the first order and take places where two surfaces of Gibbs free energy intersect, known as coexistence line in $P - T$ diagrams. At any point on this line, The following equations hold true between the two phases mentioned,
\begin{equation}
{G_1} = {G_2},{T_1} = {T_2}
\end{equation}
where the indices 1 and 2 correspond to the two different phases of the black hole. The temperature equilibrium of these two phases denotes the isothermal phase transition. We plot equation of pressure with respect to temperature for some values of $m$ and $a$ parameters and we see the effect of changing these parameters on $P - T$ diagrams in Figure 8. In this Figure, $p$ and $\tau $ are the dimensionless quantities related to pressure and temperature respectively.\\
\begin{figure}[ht]
    \centering
\subfloat[a]{\includegraphics[width=4.5cm]{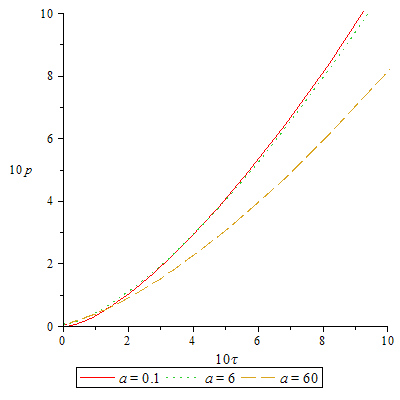}}
\qquad
\subfloat[b]{\includegraphics[width=4.5cm]{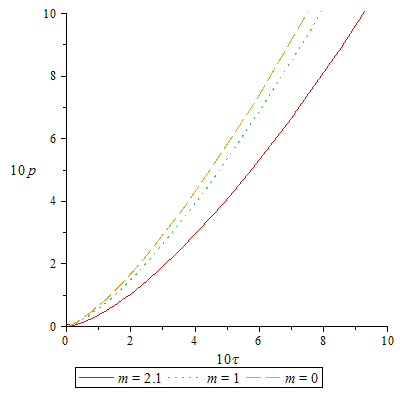}}
\caption{ $P- T$  for $k=1$, ${c_0} = 1$, ${c_1} = -2$, ${c_2} = 4$, ${c_3} = -4$ ; (a): $m = 2.1$ and $a = \{ 0.1,6,60\} $, (b): $a = 1$ and $m = \{ 2.1,1,0\}$.}
\end{figure}
In the following, we use $T - P$ diagrams to search the behavior of the thermodynamic system. This method showes a process known as Joule-Thomson expansion, which displays the change in system temperature relative to pressure at a constant enthalpy. Accordingly, we will have an isenthalpic process that can indicate heating and cooling phases. As can be seen from the diagrams in Figure 9, our process has only one cooling phase and never enters a heating phase as seen in \cite{Okcu:2016tgt} which follows a heating-cooling process in a Joule-Thomson expansion.\\  
\begin{figure}[ht]
    \centering
\subfloat[a]{\includegraphics[width=4.5cm]{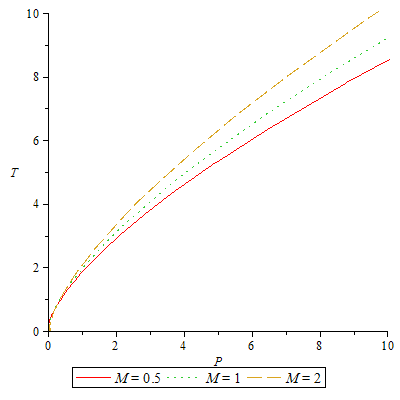}}
\qquad
\subfloat[b]{\includegraphics[width=4.5cm]{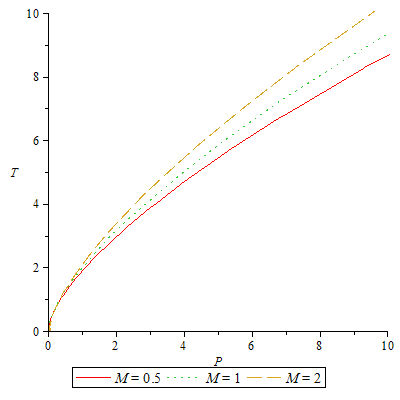}}
\caption{ $T- P$  for $k=1$, ${c_0} = 1$, ${c_1} = -2$, ${c_2} = 4$, ${c_3} = -4$, $m = 2.1$ ; (a): $a = 0.1$ and (b): $a = 1$.}
\end{figure}
Finally, we investigate the behavior of physical quantities near the critical point. In order to calculate the critical exponents discerning the behavior of physical quantities in the neighborhood of the critical point, one can use rescaled quantities $\nu  = {\upsilon  \over {{\upsilon _c}}},\tau  = {T \over {{T_c}}}$ and $ p = {P \over {{P_c}}}$ where $\upsilon  = {4 \over 3}{r_0}$ is specific volume. So, the equation of state (36) as follows
\begin{equation}
p = \left( {\frac{{{T_c}}}{{{\upsilon _c}{P_c}}}} \right)\frac{\tau }{\nu } - \left( {\frac{{3{m^2}{c_0}{c_1}}}{{8\pi {\upsilon _c}{P_c}}}} \right)\frac{1}{\nu } - \left( {\frac{{2(k + {m^2}c_0^2{c_2})}}{{3\pi \upsilon _c^2{P_c}}}} \right)\frac{1}{{{\nu ^2}}} + \left( {\frac{{4(2a - 3{m^2}c_0^3{c_3})}}{{27\pi \upsilon _c^3{P_c}}}} \right)\frac{1}{{{\nu ^3}}},
\end{equation}
where is called as law of corresponding state. We can now seek the thermodynamical behavior of the system near the critical points by redefining parameters $t,\omega $ 
\begin{equation}\nonumber
\tau  = 1 + t, \nu  = 1 + \omega.
\end{equation}
 $\nu ,\tau$ and $p$ parameters are expanded around one, so the law of corresponding state would be approximated as
\begin{equation}
p = 1 + \Theta t + \Phi t\omega  + \Omega {\omega ^3} + ...
\end{equation}
where
 \begin{align}\label{14}
&\Theta  = \frac{{{T_c}}}{{{P_c}{\upsilon _c}}},\nonumber\\
&\Phi  =  - \frac{{{T_c}}}{{{P_c}{\upsilon _c}}},\nonumber\\
&\Omega  = \frac{{72(k + {m^2}c_0^2{c_2}){\upsilon _c} + 27(\frac{3}{8}{m^2}{c_0}{c_1} - \pi {T_c})\upsilon _c^2 - 40(2a - 3{m^2}c_0^3{c_3})}}{{27\pi {P_c}\upsilon _c^3}}.
 \end{align}
To examine the behavior of the system near the critical points, one can introduce the critical exponents as \cite{Lee:2014dha}
 \begin{align}\label{15}
{C_\upsilon } = T\frac{{\partial S}}{{\partial T}}\left| {_\upsilon } \right. \propto {\left| t \right|^{ - \alpha }},\nonumber\\
\eta  = {\upsilon _l} - {\upsilon _s} \propto {\left| t \right|^\beta },\nonumber\\
{\kappa _T} =  - \frac{1}{\upsilon }\frac{{\partial \upsilon }}{{\partial P}}\left| {_T} \right. \propto {\left| t \right|^{ - \gamma }},\nonumber\\
\left| {P - {P_c}} \right| \propto {\left| {\upsilon  - {\upsilon _c}} \right|^\delta }.
 \end{align}
As is clear from the above definitions, the exponents $\alpha ,\beta ,\gamma $, and $\delta $ describe the behavior of specific heat with fixed volume, the order parameter $\eta $, the isothermal compressibility coefficient ${\kappa _T}$, and the critical isotherm, respectively. The subscripts $l$ and $s$ represent the large black hole and the small black hole, respectively, in the phase transition process.\\
The entropy $S$ does not depend on the Hawking temperature $T$, so the specific heat at constant volume ${C_\upsilon }$ is equal to zero, accordingly the corresponding critical exponent vanishes ($\alpha  = 0$). To calculate the second exponent $\beta$, one can evaluate ${\upsilon _l}$ and ${\upsilon _s}$ to find the order parameter. During the phase transition the pressure of the black hole holds changeless. It causes that the large black hole pressure equals the small black hole pressure, ${p_l} = {p_s}$ for which
\begin{equation}
1 + \Theta t + \Phi t{\omega _l} + \Omega \omega _l^3 = 1 + \Theta t + \Phi t{\omega _s} + \Omega \omega _s^3.
\end{equation}
On the other hand, from the Maxwell's equal area law, one can further obtain
\begin{equation}
\int_{{\omega _l}}^{{\omega _s}} {\omega \frac{{dp}}{{d\omega }}} d\omega  = 0  \to  \Phi t(\omega _l^2 - \omega _s^2) + \frac{3}{2}\Omega (\omega _l^4 - \omega _s^4) = 0.
\end{equation}
With two  above Eqs, one can gain
\begin{equation}
{\omega _l} =  - {\omega _s} = \sqrt {\frac{{ - \Phi t}}{\Omega }}. 
\end{equation}
So the order parameter can be obtained as
\begin{equation}
\eta  = {\upsilon _l} - {\upsilon _s} = {\upsilon _c}({\omega _l} - {\omega _s}) = 2{\upsilon _c}{\omega _l} \propto \sqrt { - t}, 
\end{equation}
 where this leads to the conclusion that $\beta  = {1 \over 2}$.\\
 The isothermal compressibility can be calculated as follows
\begin{equation}
{\kappa _T} =  - \frac{1}{{{\upsilon _c}(1 + \omega )}}\frac{{\partial \upsilon }}{{\partial \omega }}\frac{{\partial \omega }}{{\partial P}}\left| {_T} \right. \propto  - \frac{1}{{\frac{{\partial p}}{{\partial \omega }}}}\left| {_{\omega  = 0}} \right. =  - \frac{1}{{\Phi t}}.
\end{equation}
From this one can deduce that $\gamma  = 1$. The critical isotherm is an isothermal process at critical temperature $T = {T_c}$ or $t = 0$. Then we can obtain $p - 1 = \Omega {\omega ^3}$ that results $\delta  = 3$. We see that, the values of critical exponents are independent of massive and cloud string parameters.  The critical exponents in our model are the same as those mentioned in other papers \cite{Lee:2014dha}-\cite{Li:2014ixn}, and all of the models reviewed have the same scaling laws.\\
To check the behavior of temperature with respect to the radius one can plot $T - {r_0}$  diagrams  (see Figure 10) for string cloud parameter $a$ less or greater than critical value ${a_c}$ where
\begin{equation}
{a_c} = \sqrt {\frac{{ - {{(k + {m^2}c_0^2{c_2})}^3}}}{\Lambda }}  + \frac{3}{2}{m^2}c_0^3{c_3}.
\end{equation}
For $a < {a_c}$, there is a black hole from zero to one critical temperature and as the black hole grows  its temperature rises. From this critical temperature to a definite amount of temperature, we have three black holes, which, depending on the size of these three black holes, we call them small, medium and large. Then, as the temperature rises, two smaller black holes disappear and  only the large black hole will be at high temperatures.
For $a > {a_c}$ and ${c_3}<0$ at any given temperature, we have only one black hole.\\
\begin{figure}[t]
    \centering
\subfloat[A]{\includegraphics[width=5cm]{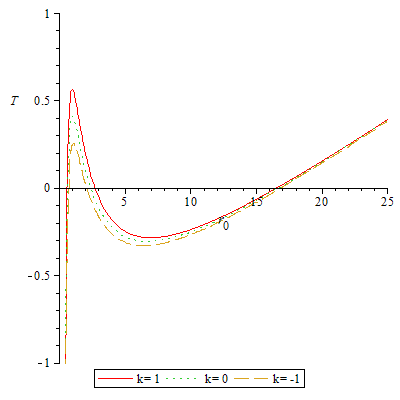}}
\qquad
\subfloat[B]{\includegraphics[width=5cm]{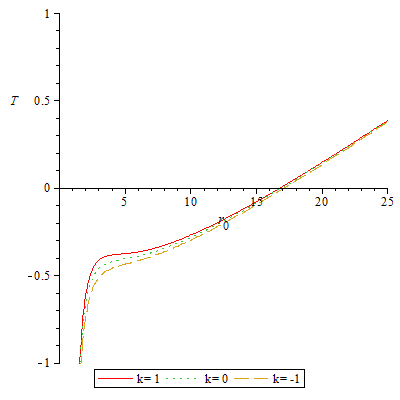}}
\qquad
\subfloat[C]{\includegraphics[width=5cm]{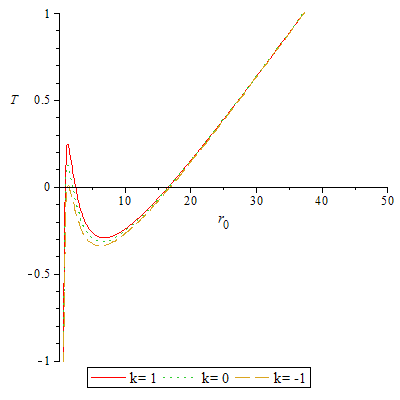}}
\caption{ $T-{r_0}$  for $d=5$, $\Lambda  = -1$, ${c_0} = 1$, ${c_1} = -2$, ${c_2} = 4$, ${c_3} = -4$ and $m = 2.1$, $(a): a=0.1< {a_c}$ =54.01, $(b): a=60 > {a_c}$ =54.01 and $(c): a=60>{a_c}$ =54.01, ${c_3} = 4$. }
\end{figure}
The specific heat  in canonical ensemble for $d = 5$ is deduced
\begin{equation}
C = \frac{{\pi [3kr_0^4 - ar_0^3 - \Lambda r_0^6 + {m^2}(\frac{{3{c_0}{c_1}}}{8}r_0^5 + \frac{{c_0^2{c_2}}}{2}r_0^4 + \frac{{c_0^3{c_3}}}{4}r_0^3)]}}{{ - 3k{r_0} + 2a - \Lambda r_0^3 - 3{m^2}c_0^2{c_2}{r_0} - 3{m^2}c_0^3{c_3}}}.
\end{equation}
The conditions of the stability of black holes are obvious from $C-{r_0}$ diagrams(see Figure 11).\\
\begin{figure}[t]
    \centering
\subfloat[A]{\includegraphics[width=5cm]{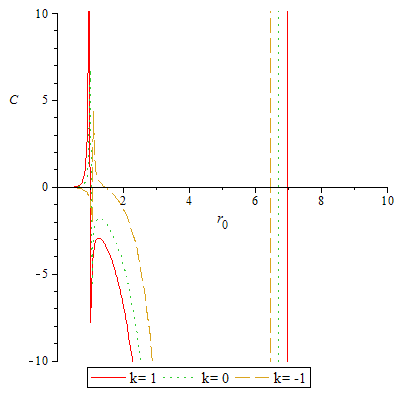}}
\qquad
\subfloat[B]{\includegraphics[width=5cm]{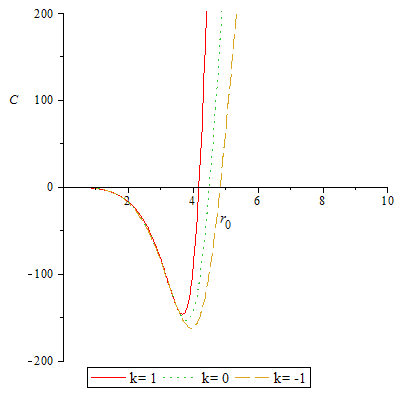}}
\qquad
\subfloat[C]{\includegraphics[width=5cm]{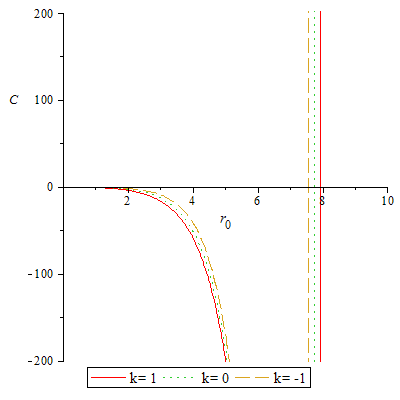}}
\qquad
\subfloat[D]{\includegraphics[width=5cm]{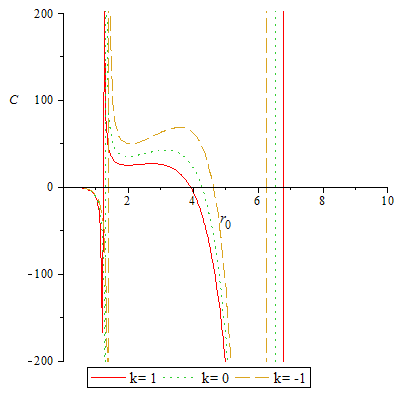}}
\caption{ $C-{r_0}$  for $d=5$, $\Lambda  = -1$, ${c_0} = 1$, ${c_1} = -2$, ${c_2} = 4$, ${c_3} = -4$, and $m = 2.1$, (a) : $a$=0.1, (b): $a$=60, (c):  $a$=0.1, ${c_3} = 4$ and (d): $a$=60, ${c_3} = 4$.}
\end{figure}
The Gibbs free energy for $d = 5$  is as follows
\begin{equation}
G =  - \frac{1}{{18}}(2a - 3{m^2}c_0^3{c_3}){r_0} + \frac{1}{{12}}(k + {m^2}c_0^2{c_2})r_0^2 - \frac{\pi }{9}Pr_0^4,
\end{equation}
which has the following critical value
\begin{equation}
{G_c} = \frac{{ - {{(2a - 3{m^2}c_0^3{c_3})}^2}}}{{96(k + {m^2}c_0^2{c_2})}}.
\end{equation}
The second order phase transition for $a < {a_c}$, is clearly evident from the $G - T$ diagrams(see Figure 12).\\
Thereafter, the presence of the massive coefficient ${c_1}$ in the pressure and temperature formulas and its absence in Gibbs free energy allows us to examine the degeneracy of states in extremal black holes. An extremal black hole is the smallest possible black hole that can exist while rotating at a given fixed constant speed. It has been suggested by Sean Carroll \cite{Carroll:2009maa} that the entropy of an extremal black hole is equal to zero. If we set $S = 0$ in Gibbs free energy formula then we obtian $G = E$. Therefore, by changing the coefficient ${c_1}$ , the pressure and temperature change in each state, but the energies of the states are the same and we get the degeneracy between states. Although the application of our proposal is unclear, but taking the extremal limit of our black hole make it possible to specify different phases with different temperatures and pressures, without any modification in the energy of different phases.\\
\begin{figure}[ht]
    \centering
\subfloat[A]{\includegraphics[width=5cm]{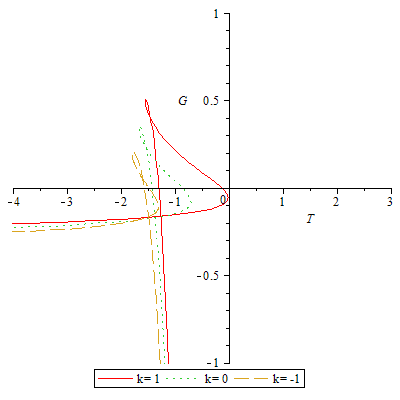}}
\qquad
\subfloat[B]{\includegraphics[width=5cm]{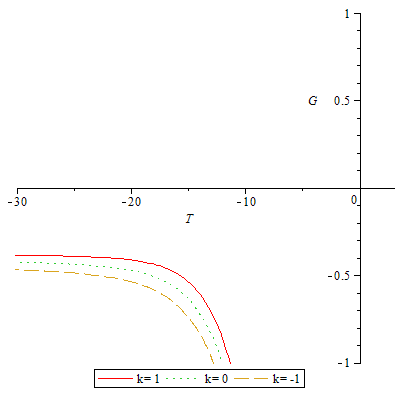}}
\caption{ $G- T$  for $d=5$, $P=0.17$, ${c_0} = 1$, ${c_1} = -2$, ${c_2} = 4$, ${c_3} = -4$,  and $m = 2.1$, (a): $a$=1$< {a_c}$  and (b): $a$=60$> {a_c}$.}
\end{figure}

\section{black hole as heat engine}
After considering black hole as thermodynamic system, one can consider such system as a heat engine. That is, black hole generates mechanical work by burning some substance. The substance used by such engine as fuel follow the equation of state. An interesting advantage of examining the thermodynamics of the black holes in the extended phase space is that the mechanical term $PdV$ in the first law provides the possibility of calculating the efficiency of these heat engines. Carnot showed that the heat engine that operates a reversible cycle  composed of two isothermals and two adiabatics, has the highest efficiency compared with the other heat engines. The efficiency of a heat engine that working between two reservoirs of temperature is given by
\begin{equation}
\eta  = \frac{W}{{{Q_H}}} = \frac{{{Q_H} - {Q_C}}}{{{Q_H}}},
\end{equation}
where ${Q_H}$,  ${Q_C}$ and  ${W}$ stand for input heat to the system, output heat from the system and mechanical work done by the system, respectively. Apart from the Carnot cycle, which is the simplest cycle that can be considered, it is difficult to find an analytical formula for the efficiency of other cycles. The first time Johnson \cite{Johnson:2014yja} proposed a simple cycle in $P - V$ plane, known as squared cycle constructed by two adiabatics/isochorics and two isobarics as shown in Figure 13. Along the isobar $1 \to 2$ and $3 \to 4$, system produces work so that the total work done by this heat engine is equal to the area of the rectangle $1 \to 2 \to 3 \to 4$. Also the input heat to the system along the isobar  $1 \to 2$ and the output heat from the system along the isobar $3 \to 4$ is obtained  from the first law of thermodynamics
\begin{equation}
dH = \delta Q + VdP.
\end{equation}
Since the pressure is constant during the isobaric process it results that the heat absorbed/emitted  is equal to the difference between the enthalpy of the two points at the beginning and the end of the process so the efficiency is as follows
\begin{equation}
\eta  = 1 - \frac{{{M_3} - {M_4}}}{{{M_2} - {M_1}}}.
\end{equation}
For the case of our black hole we plot the diagrams for efficiency according to the string cloud parameter $a$ in four and five dimensions (see Figure 14). We see that the efficiency increases by rising the string cloud  parameter.  On the other hand, given that the maximum value of efficiency is less than one, an upper limit on the string cloud parameter can be applied. Also, the gradient of the graph in four dimensions is more than five dimensions.
\begin{figure}[t]
    \centering
\subfloat[A]{\includegraphics[width=4.5cm]{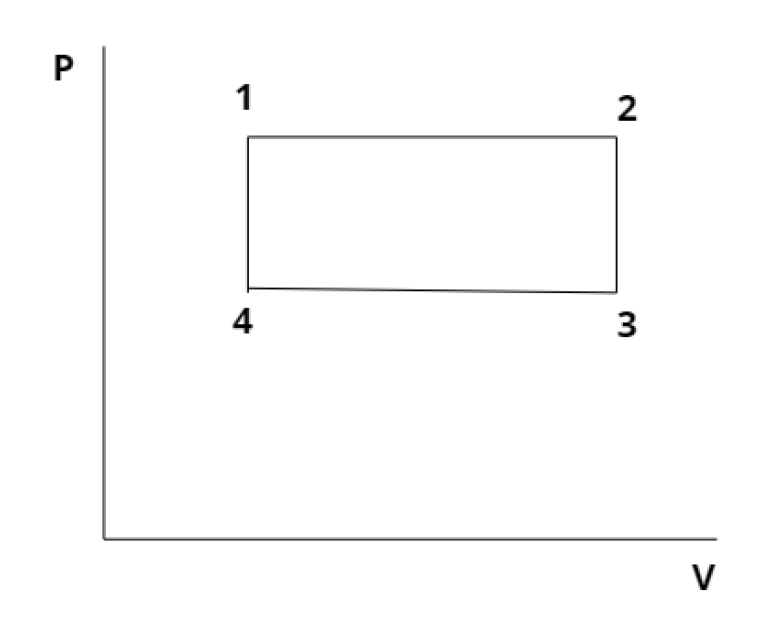}}
\caption{ A square cycle in the $ P-V$ plane.}
\end{figure}
\begin{figure}[t]
    \centering
\subfloat[A]{\includegraphics[width=5cm]{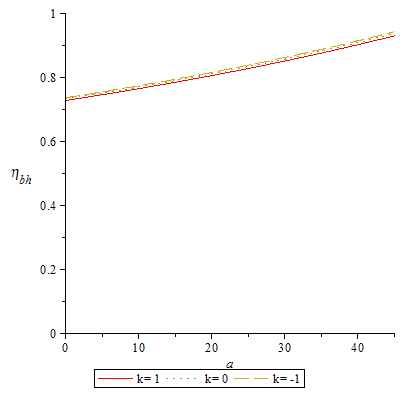}}
\qquad
\subfloat[B]{\includegraphics[width=5cm]{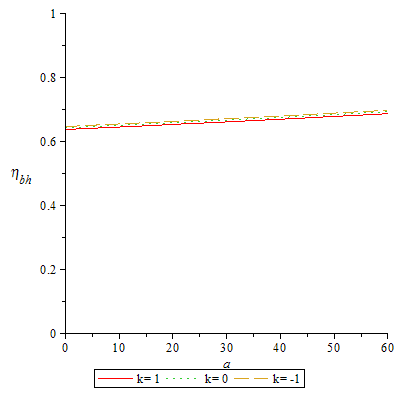}}
\caption{ $\eta  - a$  for  ${S_1}={S_4} = \pi $,  ${S_2}={S_3} = 3\pi $, ${P_1}={P_2} = 4 $, ${P_3}={P_4} =1 $,  ${c_0} = 1$, ${c_1} =  - 2$, ${c_2} = 4$, ${c_3} = 4$,  and $m = 2.1$, (a): $d=4$ and (b): $d=5$.}
\end{figure}
 \section{Conclusion}
\noindent In this paper we observed that the massive gravity minimally coupled to a cloud of strings can not exhibit Van der Waals like behavior for AdS--Schwarzschild black hole in four dimensions, but in five dimensions this modification takes place. This behavior can be attributed to the absence of one term from the mass episode namely the third massive term in the equation of  state in four dimensions. But in five dimensions, the massive parameter and the string cloud parameter, the latter only in flat topology, play a decisive role in the Van der Waals like behavior.\\
 It is interesting to note that the string cloud term in each dimensions treats like the final massive term. Also the string cloud parameter affects the number of event horizons and the quality of the black holes stability, so we found a critical value for this parameter. We concluded that for values smaller than the critical value of the string cloud parameter, there is a Van der Waals-like behavior but not for values larger than that.\\
We observed the critical behavior of the system for temperatures below the critical temperature by plotting $P -{r_0}$ diagrams. We also investigated the critical behavior of the system by drawing coexistence lines on the $P - T$ plane. In the following, from the $T - P$ figures, we founnd that the Joule-Thomson effect does not occur in our model.  Then, we saw that the critical exponents in our model are the same as the other models.\\
Finally we saw that the string cloud parameter affects  the efficiency of the black hole heat engine  and we showed for squared cycle, the efficiency grows up as the string cloud parameter increases and approaches to one. The upper limit that can be considered for a string cloud parameter becomes larger by increasing the dimensions.


\end{document}